# Near-filed SAR Image Restoration with Deep Learning Inverse Technique: A Preliminary Study

Xu Zhan, Xiaoling Zhang, Wensi Zhang, Jun Shi, Shunjun Wei, Tianjiao Zeng
University of Electronic Science and Technology of China
Chengdu, China
Email: {zhanxu, zws@std.uestc.edu.cn}, {xlzhang, shijun, weishunjun, tzeng@uestc.edu.cn}

*Abstract*—Benefiting from a relatively larger aperture's angle, and in combination with a wide transmitting bandwidth, near-field synthetic aperture radar (SAR) provides a high-resolution image of a target's scattering distribution-hot spots. Meanwhile, imaging result suffers inevitable degradation from sidelobes, clutters, and noises, hindering the information retrieval of the target. To restore the image, current methods make simplified assumptions; for example, the point spread function (PSF) is spatially consistent, the target consists of sparse point scatters, etc. Thus, they achieve limited restoration performance in terms of the target's shape, especially for complex targets. To address these issues, a preliminary study is conducted on restoration with the recent promising deep learning inverse technique in this work. We reformulate the degradation model into a spatially variable complex-convolution model, where the near-field SAR's system response is considered. Adhering to it, a model-based deep learning network is designed to restore the image. A simulated degraded image dataset from multiple complex target models is constructed to validate the network. All the images are formulated using the electromagnetic simulation tool. Experiments on the dataset reveal their effectiveness. Compared with current methods, superior performance is achieved regarding the target's shape and energy estimation.

## I. INTRODUCTION

Near-field synthetic aperture radar (SAR) can obtain a high-resolution target image. It can help with scattering information retrieval in the application scenarios like automatic target recognition (ATR) [1]. It can also be a useful verification or diagnosis tool for military stealth equipment during their stealth design and routine maintenance phases [2]. However, images inevitably suffer degradation from side effects, including side lobes, clutters, and noises. These factors directly result in the target's shape loss, greatly hinder the image quality, and negatively impact recognition and detection.

To restore the image, in other words, invert the degradation, current methods take different methodologies. Considering the sparsity of main scatters in a target [2], a sparsity-based method is proposed in [3]. However, only strong scatters are restored, and those relatively weak scatters are lost. While weak scatters are generally distributed scatters that make up the shape, they matter for complex targets like aircraft.

Taking another methodology that considers the point spread function (PSF), greedy methods are proposed, like CLEAN [4]. Still, it is suitable for simple targets but not complex targets with dense scatter distribution. Cumulative errors during the greedy-solving process are not tolerable. Besides the methods above, nonlinear-filtering methods like the popular spatially variant apodization (SVA) [5] exist. exist. However, as they assume the PSF is static over the scene or the PSF is orthogonal in two-dimensional space, they are not fit-ted well for near- field SAR, where the PSF is neither orthogonal nor static.

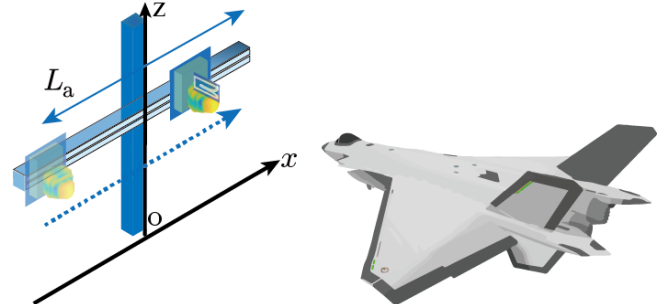

Fig. 1. Typical near-field SAR imaging mode illustration.

In our previous work [6], we take a new methodology, considering the PSF's spatially variation and the non-orthogonality. A two-dimensional spatial-variant degradation model is established. The image is then restored by deconvolution. However, the process only considers the target's sparse feature. This makes it suffer performance loss faced with complex targets. We believe that more sufficient features are needed for more complex targets. Besides, the solving process needs hyper-parameter tuning, which is time-consuming and not easy.

Deep learning has shown prominent performance in the radar community in recent years. The most important reason is its strong ability to extract features we need for our problem. Among the existing research, a promising trend, using deep learning to solve inverse problems [7], [8], is closely related to our problem. Like many other problems in the area, e.g., imaging, super-solution, despeckling, [9] the restoration problem can be attributed to an inverse problem. In these other problems, utilizing such a technique has shown superior performance [7], [8], [10].

With these considerations and observations, a preliminary study is conducted in this work for the near-field SAR image restoration problem. Starting from the established spatially variable complex convolution degradation model, we firstly review the traditional proximal-gradient-descend [11] based solver for restoration, which consists of multiple iteration blocks. For each block, they share similar sub-steps. With this observation, inspired by deep-unfolding [12] , a complex convolutional neural network is designed to mimic this process. And instead of the insufficiently hand-crafted sparse feature that we adopted before [6], an encoder-decoder network is embedded for learning more sufficient features.

To study the performance of these two networks, instead of simple point targets, we establish a more truthful complex target dataset utilizing the electromagnetic simulation software. Results on the dataset show promising performances in aspects like target shape restoration and scattering energy restoration. More details are presented in the rest of the paper.

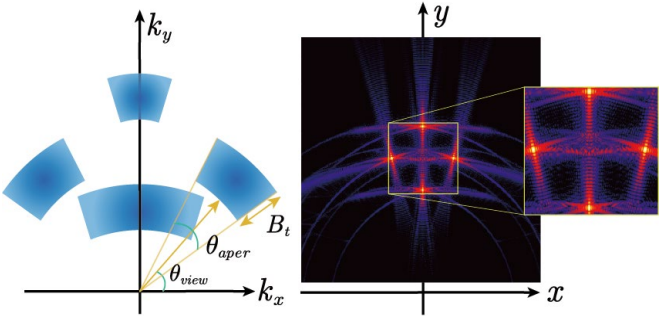

Fig. 2. Near-field SAR 2D spatially variant PSF responses. Left: responses in the wavenumber domain. Right: responses in the spatial image domain.

## II. SPATIALLY VARIABLE COMPLEX-CONVOLUTION DEGRADATION MODEL

A target's SAR image is the convolution of PSF responses and the target's true image. The PSF response of a scatter is dependent on the imaging modality of the system. The imaging mode of near-field SAR is shown in Fig. 1. Different from the far-field SAR, the target is at a closer range. From the perspective of the wavenumber domain-K space, a scatter's PSF is dependent on its spatial spectrum [14]. This spectrum is an annular sector in the wave domain, whose orientation, extended angle, and side length are determined by the angle of view, the angle of synthetic aperture, and the transmitting bandwidth, respectively. And the former two factors are spatially variant in the two-dimensional (2D) directions in the near-field SAR. An example in shown in Fig. 2.

Because the synthetic aperture length is fixed, as seen in Fig. 2, the closer scatter has a relatively larger synthetic angle, which results in larger bandwidth and higher resolution in the azimuth direction. In other words, the resolution is spatially variant. The resolution formulas can also verify this. For the scatters that have the view-angle of 90 degrees, their resolutions can be expressed as

$$d_x = \frac{\lambda R}{2L}, \quad d_y = \frac{c}{2B} \quad (1)$$

where $d_x$ and $d_y$ are the resolution at the azimuth and range direction, respectively. $L$ is the aperture length. $R$ is the range. $\lambda$ is the wavelength of the center frequency. $c$ is the light speed. $B$ is the transmitting bandwidth. It can be seen that the resolution at the azimuth direction is proportional to the range. This can be observed from the enlarged part within a yellow bounding box at the right of Fig. 2.

Besides variety in this aspect, the resolution directions-the sidelobe directions are also spatially variant. The sidelobes are due to limited bandwidth observation. Their directions are approximately orthogonal to the directions of the annular sector' sides [15]. As seen in the right part of Fig. 2, the two scatters in the middle have different sidelobe directions compared to the other two scatters' sidelobes. Notably, all the scatters' sidelobes are non-orthogonal. In other words, the sidelobes are coupled in two directions. At the azimuth direction, the sidelobe even divides into two curved paths. The closer to the system, the more apparent the division is, and the more curved the sidelobes are. For complex targets, which are normally large-scale at the degree of the aperture's length, due to the complex situation of the PSFs, the degradation would severely cause shape loss. An example of an aircraft model is shown in Fig. 3. Compared with the actual scattering distribution, although main strong point-type scatters are reserved, their resolution is worse due to the limited bandwidth observation. And their spa-tial locations suffer deviations due to the coherent summation of PSFs. What's worse, weak distributed-type What's worse is that weak distributed-type scatters are missing or interfere with each other, resulting in the shape of the target being lost, including the geometry and structure.

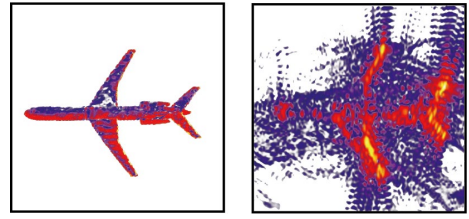

Fig. 3. Degradation of an aircraft model image. Left: calculated scattering distribution true image. Right: Near-field SAR image with a serve degradation.

Next, we characterize the degradation process mathematically. The degraded image of one scatter can be considered as the complex convolution of its true image and the corresponding PSF as follows.

$$\mathbf{Yd}(x,y) = a_0 \boldsymbol{\delta}(x-x_0, y-y_0) * \mathbf{H}(x,y) \quad (2)$$

Where $a_0 \boldsymbol{\delta}(x-x_0, y-y_0)$ and $\mathbf{H}(x,y)$ are the true image and the PSF image, respectively. The true image is a 2D ideal impulse function where the impulse is located at the scatter position $(x_0, y_0)$ with its amplitude being the scattering coefficient $a_0$. Thus, for the degraded image of the whole target, the spatially variable complex-convolution degradation model can be expressed as follows.

$$\mathbf{Y}(x,y) = \sum_i \mathbf{H}_i(x,y) * a_i \boldsymbol{\delta}(x-x_i, y-y_i) + \mathbf{N}(x,y) \quad (3)$$

Where $\mathbf{H}_i(x,y)$ and $a_i \boldsymbol{\delta}(x-x_i, y-y_i)$ are the $i$th scatter's true image and PSF image, respectively. Here, we consider the practical existence of clutter and noise $\mathbf{N}(x,y)$. Utilizing the relationship of PSF and the corresponding spatial spectrum and with some modifications, the model can be expressed more compactly as follows.

$$\mathbf{Y} = f_2^\dagger \left( f_1^\dagger \left( \mathbf{H}_f \cdot f_1(f_2(\mathbf{X})) \right) \right) + \mathbf{N} \quad (4.1)$$

$$\mathbf{H}_f = [\cdots \quad \boldsymbol{m}_f^i \circ \boldsymbol{h}_f^i \quad \cdots] \quad (4.2)$$

$$\boldsymbol{h}_f^i = [\ldots \quad e^{-i(k_x x_i + k_y y_i)} \quad \ldots]^\mathrm{T} \quad (4.2)$$

Where $\mathbf{X}$, $\mathbf{Y}$ and $\mathbf{N}$ are true, degraded and noise-clutter images, respectively. $f_1$ is the folding operator that modifies a 2D image into its column-wise vector form [16], and $f_1^\dagger$ is the corresponding adjoint operator. $f_2$ is the 2D Fourier transform operator and $f_2^\dagger$ is the corresponding adjoint operator. $\circ$ denotes the Hadamard product. $[\cdot]^\mathrm{T}$ denotes matrix/vector transpose. $\boldsymbol{m}_f^i$ is the vectorized mask of $i$th PSF in the wavenumber domain, where its element value is one within the observed spectrum and zero otherwise. With this model, the restoration process can be formed as an optimization problem as follows.

$$\widehat{\mathbf{X}} = \mathrm{argmin}_\mathbf{X} \frac{1}{2} \|\mathbf{Y} - f(\mathbf{X})\|_\mathrm{F}^2 + \beta g(\mathbf{X}) \quad (5)$$

Where $f(\mathbf{X}) = f_2^\dagger \left( f_1^\dagger \left( \mathbf{H}_f \cdot f_1(f_2(\mathbf{X})) \right) \right)$, and $\|\cdot\|_\mathrm{F}^2$ denotes the matrix Frobenius norm. Here $g(\mathbf{X})$ is a regularization term that embeds the target's feature, the prior knowledge we have about the target. If the restored $\widehat{\mathbf{X}}$ has the features we expect, its value is small. Otherwise, the value is large to punish. $\beta$ is a weight for balancing.

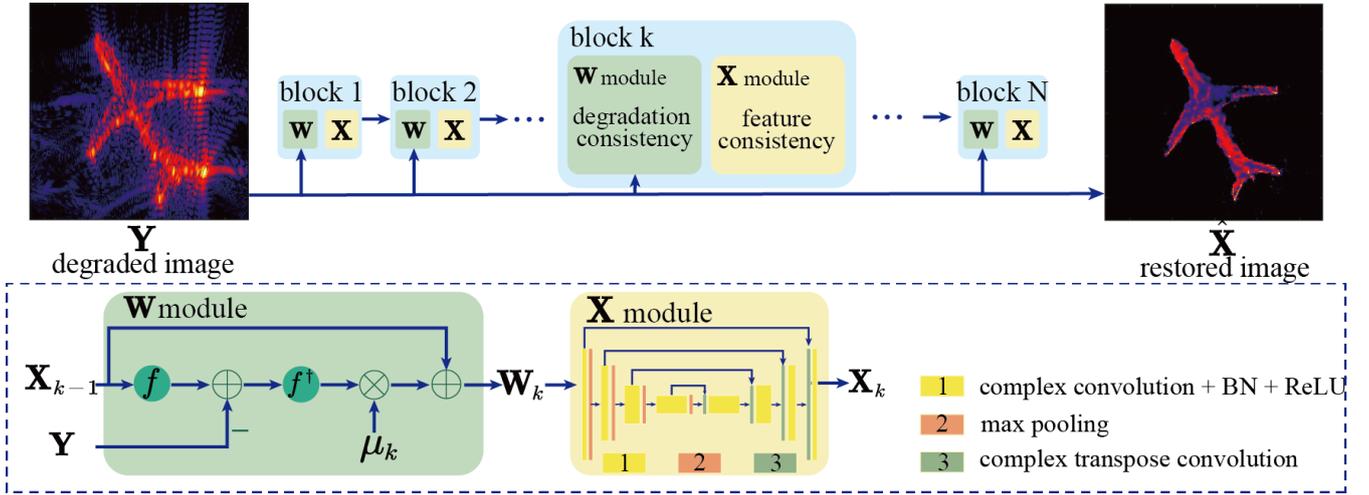

Fig. 4. The overall structure of the designed model-based restoration network. The network contains cascaded basic blocks. Each block has the same structure, including two modules to perform corresponding steps in (6).

## III. MODEL-BASED RESTORATION

In this section, we review the traditional method to solve the problem in (5), then design two methods with the deep learning inverse technique.

### A. Traditional Restoration with Iteratively Solving

The problem in (5) can be solved with the proximal gradient algorithm [11], which iteratively solves the problem. Within each iteration block, two steps are included as follows.

$$\mathbf{W}_{k+1} = \mathbf{X}_k + \mu f^\dagger(\mathbf{Y} - f(\mathbf{X}_k)) \quad (6.1)$$

$$\mathbf{X}_{k+1} = prox(\mathbf{W}_{k+1}; g, \beta) \quad (6.2)$$

Where $\mathbf{X}_k$ and $\mathbf{X}_{k+1}$ are the outputs of $k$th and $(k+1)$th iteration block, respectively. $\mu$ is the iterative step size and a hyper-parameter as $\beta$ that has direction impact on the quality of the final result. $prox(\cdot)$ is a proximal operator that is depends on the regularization term $g$ and the weight $\beta$. For example, in our previous work, we adopt the regularization term be the L1 norm of the target image, thus resulting in a soft-thresholding operator [6]. $f^\dagger$ is the corresponding adjoint operator of $f$. For an arbitrary input $\mathbf{A}$, it operates as follows.

$$f(\mathbf{A}) = f_2^\dagger \left( f_1^\dagger \left( (\mathbf{H}_f^*)^\mathrm{T} \cdot f_1(f_2(\mathbf{A})) \right) \right) \quad (7)$$

Where $\left(\mathbf{H}_f^*\right)^\mathrm{T}$ denotes the conjugate transpose of $\mathbf{H}_f$. In terms of the restoration quality, the hyperparameters need to adjust to find optimal values, which is inconvenient and time-consuming. What's more important is that for complex targets, a sufficient target feature may not be easy to find manually. And even the feature's corresponding regularization is established, obtaining the proximal operator with an explicit expression may not be possible.

### B. Model-based Restoration Network

As analyzed before, the solving process contains iterative blocks, where the input is the degraded image, and the output is the restored image. If we perceive this process as a forward process from the degraded to the restored one, intuitively, we can find this process resembles the forward flow of a deep-learning network. With this observation, a network can be designed with basic blocks cascading in a serial structure. And in each block, two basic modules are utilized to mimic the steps in (6). As the model guides this design, this network inherently embeds the degradation mechanism of Near-field SAR. The overall structure of the network is shown in Fig. 4. N blocks are cascaded to form the network.

The N blocks share the same structure. Within each block, two modules are contained:

**W** *module.* This module is to mimic the step in (6.1). In the $k$th block, it includes the following calculation.

$$\mathbf{W}_k = \mathbf{X}_{k-1} + \mu_k f^\dagger(\mathbf{Y} - f(\mathbf{X}_{k-1})) \quad (8)$$

Here, $\mathbf{X}_{k-1}$ is the output of the last block. Compared to the formula in (6), $\mu_k$ here is learnable. And instead of sharing one common learnable step size, they can have different values. This variable step size mimics the traditional backtracking step-size rule [17] for a more stable optimization path. The calculation in this module is directly related to the first term in (6). As analyzed in [17], it can be perceived as a gradient descend step to fulfill the first term. Relating to our restoration problem, this step enables that the restored image and the degraded image can be related by $f$, in other words, being consistent with the degradation model.

**X** *module.* This module is to mimic the step in (6.2). In the $k$th block, it includes the following calculation.

$$\mathbf{X}_k = \mathrm{U}prox(\mathbf{W}_k; \boldsymbol{\Theta}) \quad (9)$$

Different from the one in (6.2), $\mathrm{U}prox(\cdot; \boldsymbol{\Theta})$ is an implicit proximal operator with learnable parameters $\boldsymbol{\Theta}$. This operator corresponds to the sub-network in the yellow box. Its architecture adopts the U-net style that consists of an encoder and a decoder. This type of architecture has a strong feature extraction ability verified in multiple areas. Starting from the input, the intermediate restored image $\mathbf{W}_k$ passes multiple linear convolution transform operations, revealing features in different transformed domains. Besides, multiple max-pooling operations enable down-sampling to reveal features at different levels. And nonlinear ReLU operations that follow the corresponding convolutional transform operation mimic the traditional thresholding role, filtering out the non-ideal factors from the degradation. Then, the feature images are up-sampled through transpose convolutions, symmetrical linear convolution transforms, and ReLU operations, backing to the original domain. Along with this process, features at different levels are fused through skipping connections.

Another interpretation of the whole process in the module can be obtained from the methodology of project gradient descend [18]. In the image space, the input $\mathbf{W}_k$ can be viewed as a derivation from the sub-space where the actual image is located. This step functions as a projection where the derivation one is projected back to the actual location. Thus, the restored image's feature is consistent with the actual image's.

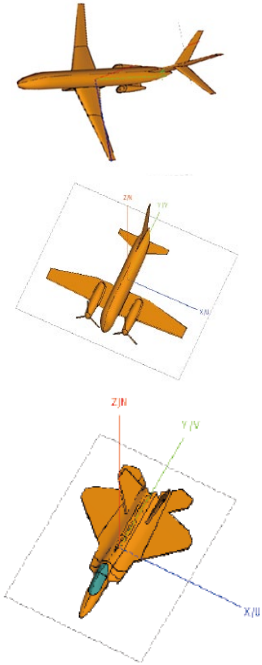

Fig. 5. Examples of aircraft models for the simulated dataset.

TABLE I. SIMULATION PARAMETERS

| Parameter | Value |
| --- | --- |
| Center frequency | 10 GHz |
| Bandwidth | 2 GHz |
| Aperture length | 2 m |
| Center range | 3 m |
| Target size | ≈2 m × 2 m |

This restoration network involves learned parameters $\Theta$ and $\mu_k, k = 1,2,...N$, which are learned through backpropagation training. The learning loss is set as MSE between the estimated clean image and the truly clean image.

## IV. EXPERIMENTS

### A. Simulated Dataset

To study the performance of two networks, we simulate a dataset containing pairs of clean and degraded images. Instead of using simple point-type scatters for the study, we choose complex airplane models as targets. Ten different types of airplane models are adopted. Three of them are illustrated in Fig. 5. With these models, we use the electromagnetic simulation software to obtain their clean images and then use the near-field SAR imaging mode in Fig. 1 to get echoes. And the time-domain correlation imaging algorithm BP [19] is adopted to obtain the final degraded images. The main simulation parameters are summarized in table 1. In total, 120 pairs of images are simulated. These images cover different airplane models from different angles of view. 100 pairs are used for training, and 20 pairs are used for testing.

### B. Implementation Details

The two networks are constructed in the Pytorch framework on a personal computer with NVIDIA GTX 3070Ti. The CPU is intel i9-9900 and the memory size is 32 GB. The Adam optimizer is applied for training by 100 epochs. The degraded image size is 256 × 256. For the first network, without any approximation, there would be 65536 different PSFs. This would bring a huge computation and storage burden. Thus, for a local area of 8 × 8, we use one common PSF to approximate. The number of iteration blocks in the network is set to four empirically. We find more blocks would not increase the restored image quality much but increase the parameters to learn and computation burden.

We compare the two networks with the sparsity-based method, CLEAN, SVA, and the deconvolution method we proposed before. Two metrics, MSE and SSIM [9] are used for quantitative comparison. MSE evaluates the scattering energy restoration accuracy, and SSIM evaluates the shape restoration ability. For MSE, a lower value is better, and for SSIM, a higher one is better.

### C. Experiment Results

An example of visual comparison result with other methods is shown in Fig. 6. Except for the method of SVA, all the other three methods can only restore the main point-type scatters as shown in Fig. 6(c), (d), and (f). Although the sidelobes are suppressed well, the weak distributed-type scatters are also lost. There is barely any shape restored. The result of SVA in Fig. 6(e) shows that this method has the worst performance. Although the resolution is enhanced a little, all the sidelobes remain.

A great contrast exists between the network-based method and traditional methods. As shown in Fig. 6(g), a much more balanced restoration performance is obtained. While suppressing all the sidelobes, the strong and weak scatters are mostly restored with a much higher resolution. The quantitative results are shown in Table 2. Inconsistent with the visual perception, the network-based methods achieve much better performance, especially for the SSIM metric. This reveals their advantages in structure and geometry restoration. A different comparison example is shown in Fig. 7. the results show that it suits the degradations of other targets. The quantitative results are shown in Table 3. These two examples reveal the generalization ability to a certain extent.

With a closer look at the partial structures in the target, like the engines, we find that the shape is not restored as well as other parts. It has the potential discontinuity problem. It may be because we only take the MSE training loss without any constraints on their structures. A possible solution is to adopt more complex and comprehensive training loss functions, e.g., perception loss, adversarial loss, etc [20].

## V. CONCLUSION

In this work, faced with the near-field SAR image degradation problem, we consider the situation of more complex targets. Thinking from the spatially variable PSF, we establish the spatially variable complex-convolution model to describe the degradation process. From the traditional restoration-solving process, with the deep learning inverse technique, a restoration network is designed and studied. The degradation model is embedded in the network's forward process or the network's training loss. A study on the simulated dataset reveals their superior performance. The shape and the scattering energy can be restored much better than traditional methods. Through this preliminary study that shows their potential ability, further research on the real data has been planned as our subsequent work.

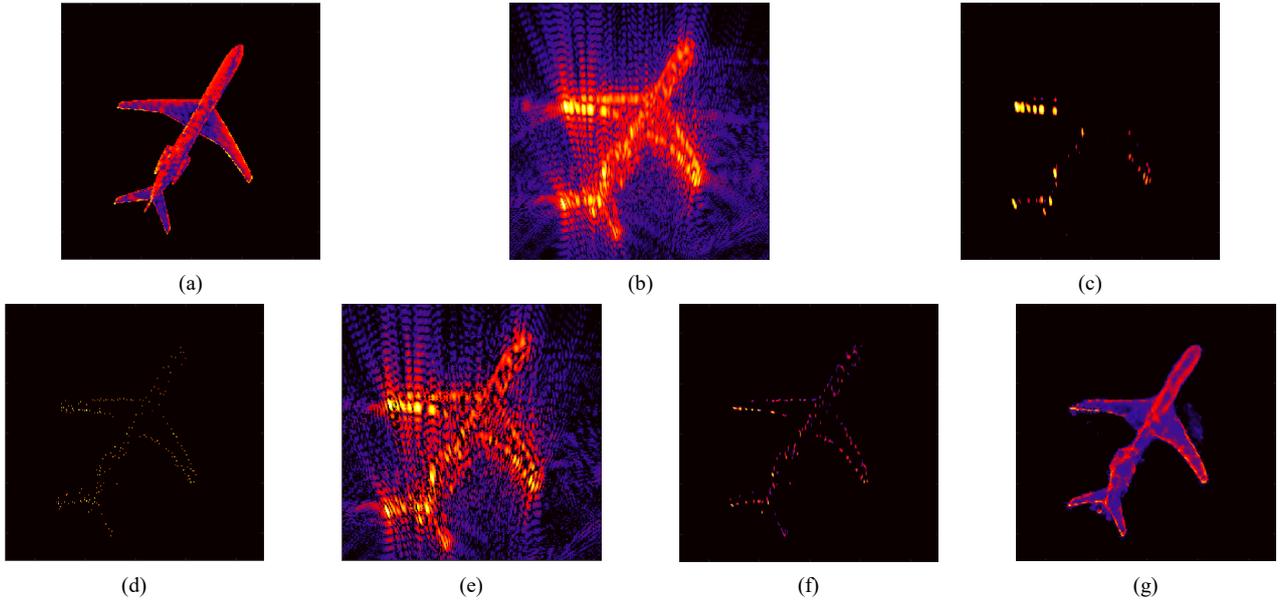

Fig. 6. Restoration results of different methods. (a) Ground-truth clean image. (b) Original degraded image. (c) Restored image by the sparsity-based method. (d) Restored image by the CLEAN method. (e) Restored image by the SVA method. (f) Restored image by the deconvolution method. (g) Restored ismage by the restoration network in Fig. 4.

TABLE II. QUANTITATIVE COMPARISON RESULTS IN FIG. 6

| Metric | sparsity-based method | CLEAN | SVA | deconvolution method | Restoration network |
|---|---|---|---|---|---|
| MSE | 0.010 | 0.004 | 0.025 | 0.003 | **0.002** |
| SSIM | 0.871 | 0.900 | 0.499 | 0.920 | **0.944** |

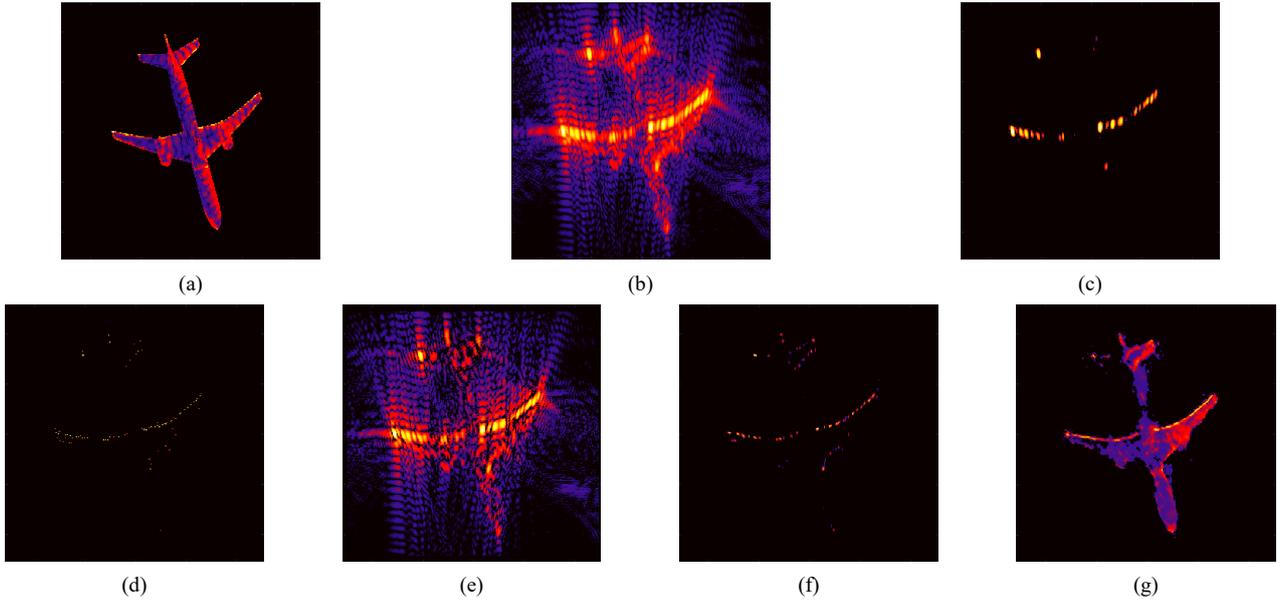

Fig. 7. Restoration results of a different model. (a) Ground-truth clean image. (b) Original degraded image. (c) Restored image by the sparsity-based method. (d) Restored image by the CLEAN method. (e) Restored image by the SVA method. (f) Restored image by the deconvolution method. (g) Restored image by the restoration network in Fig. 4.

TABLE III. QUANTITATIVE COMPARISON RESULTS IN FIG. 7

| Metric | sparsity-based method | CLEAN | SVA | deconvolution method | Restoration network |
|---|---|---|---|---|---|
| MSE | 0.008 | 0.004 | 0.014 | 0.005 | **0.002** |
| SSIM | 0.895 | 0.898 | 0.648 | 0.896 | **0.947** |